\documentclass[prl,twocolumn]{revtex4}
\usepackage{graphicx}
\usepackage{amssymb,amsmath,bm}
\usepackage{enumerate}

\newcommand{\be}{\begin{eqnarray}}
\newcommand{\ee}{\end{eqnarray}}

\begin{document}

\title{Distributed chaos in turbulent wakes}

\author{A. Bershadskii}

\affiliation{
ICAR, P.O. Box 31155, Jerusalem 91000, Israel
}

\begin{abstract}

Soft and hard spontaneous breaking of space translational symmetry (homogeneity) have been 
studied in turbulent wakes by means of distributed chaos. In the case of the soft translational symmetry breaking the vorticity correlation integral $\int_{V} \langle {\boldsymbol \omega} ({\bf x},t) \cdot  {\boldsymbol \omega} ({\bf x} + {\bf r},t) \rangle_{V}  d{\bf r}$ dominates the distributed chaos and the chaotic spectra $\exp-(k/k_{\beta})^{\beta }$  have $\beta =1/2$. In the case of the hard translational symmetry breaking, control on the distributed chaos is switched from one type of fundamental symmetry to another (in this case to Lagrangian relabeling symmetry). Due to the Noether's theorem the relabeling symmetry results in the inviscid helicity conservation and helicity correlation integral $I=\int \langle h({\bf x},t)~h({\bf x}+{\bf r}, t)\rangle d{\bf r}$ (Levich-Tsinober invariant) dominates the distributed chaos with $\beta =1/3$. Good agreement with the experimental data has been established for turbulent wakes behind a cylinder, behind grids (for normal and super-fluids) and for bubbling flows. In the last case even small concentration of bubbles leads to a drastic change of the turbulent velocity spectra due to the hard spontaneous symmetry breaking in the bubbles' wakes. 
\end{abstract}

\maketitle

\section{Soft and hard spontaneous symmetry breaking}

 The spontaneous breaking of space translational symmetry (homogeneity) has been 
studied for the first time for {\it weak} turbulence in a recent paper Ref. \cite{nrz}. For {\it strong} turbulence it was studied in Ref. \cite{b1} by means of the distributed chaos. It was shown in the Ref. \cite{b1} that in one of the possible scenarios scaling of the group velocity of the waves (pulses) driving the distributed chaos
$$
\upsilon (\kappa ) \propto |\gamma|^{1/2}~\kappa^{\alpha} \eqno{(1)}
$$
is dominated by the vorticity correlation integral
$$
\gamma = \int_{V} \langle {\boldsymbol \omega} ({\bf x},t) \cdot  {\boldsymbol \omega} ({\bf x} + {\bf r},t) \rangle_{V}  d{\bf r} \eqno{(2)}
$$
with $\alpha=1/2$ from the dimensional considerations. It results in the chaotic stretched exponential spectra
$$
E(k )  \propto \exp-(k/k_{\beta})^{\beta}  \eqno{(3)}
$$
with 
$$
\beta =\frac{2\alpha}{1+2\alpha}   \eqno{(4)}
$$ 
i.e. $\beta =1/2$. This is a soft scenario, since we are still dealing with the translational symmetry \cite{b1}.
\begin{figure}
\begin{center}
\includegraphics[width=8cm \vspace{-1cm}]{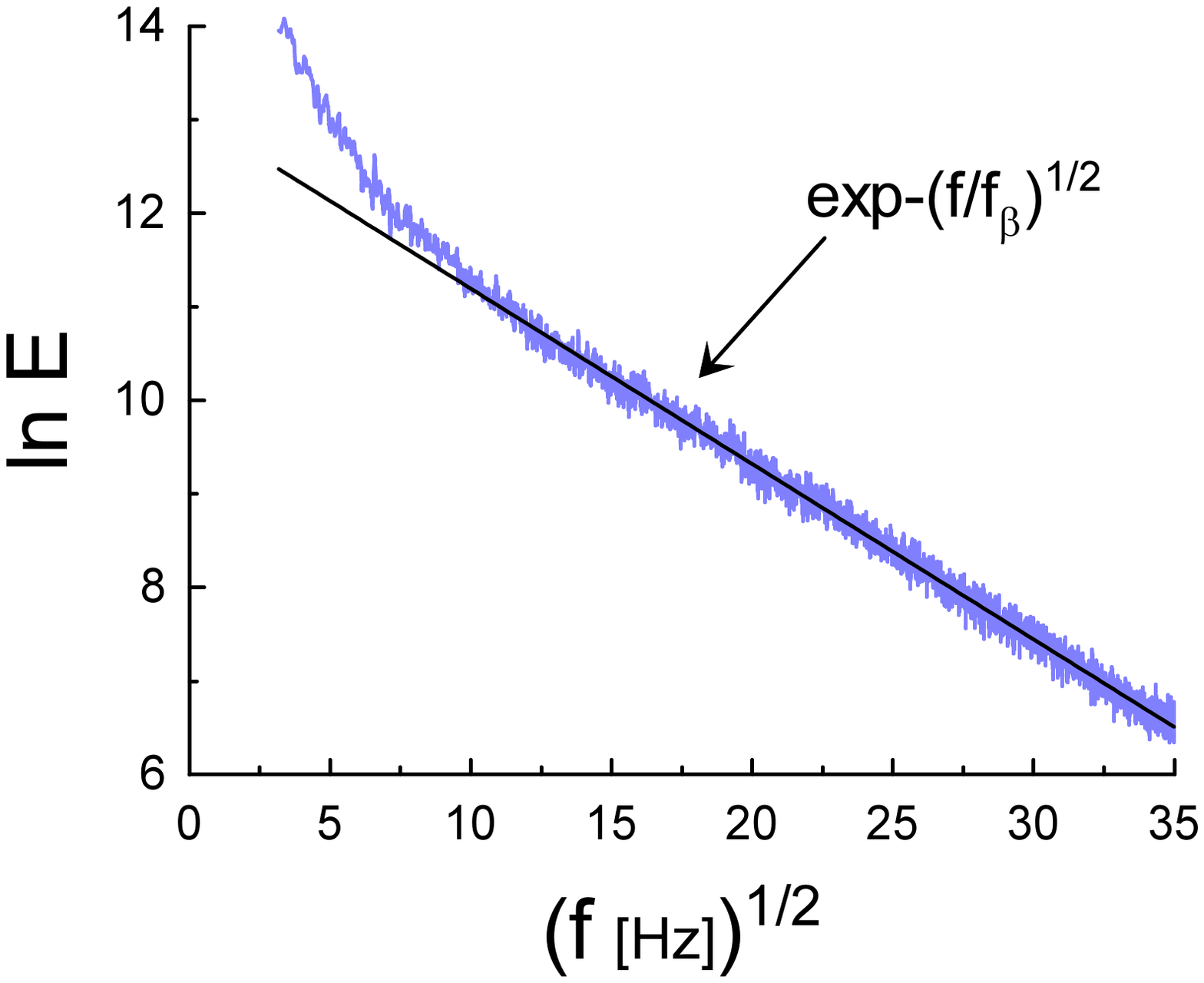}\vspace{-3.9cm}
\caption{\label{fig1} Logarithm of the frequency power spectrum of the temperature fluctuations, measured at the wake centerline, against $f^{1/2}$.}
\end{center}
\end{figure}
\begin{figure}
\begin{center}
\includegraphics[width=8cm \vspace{-1cm}]{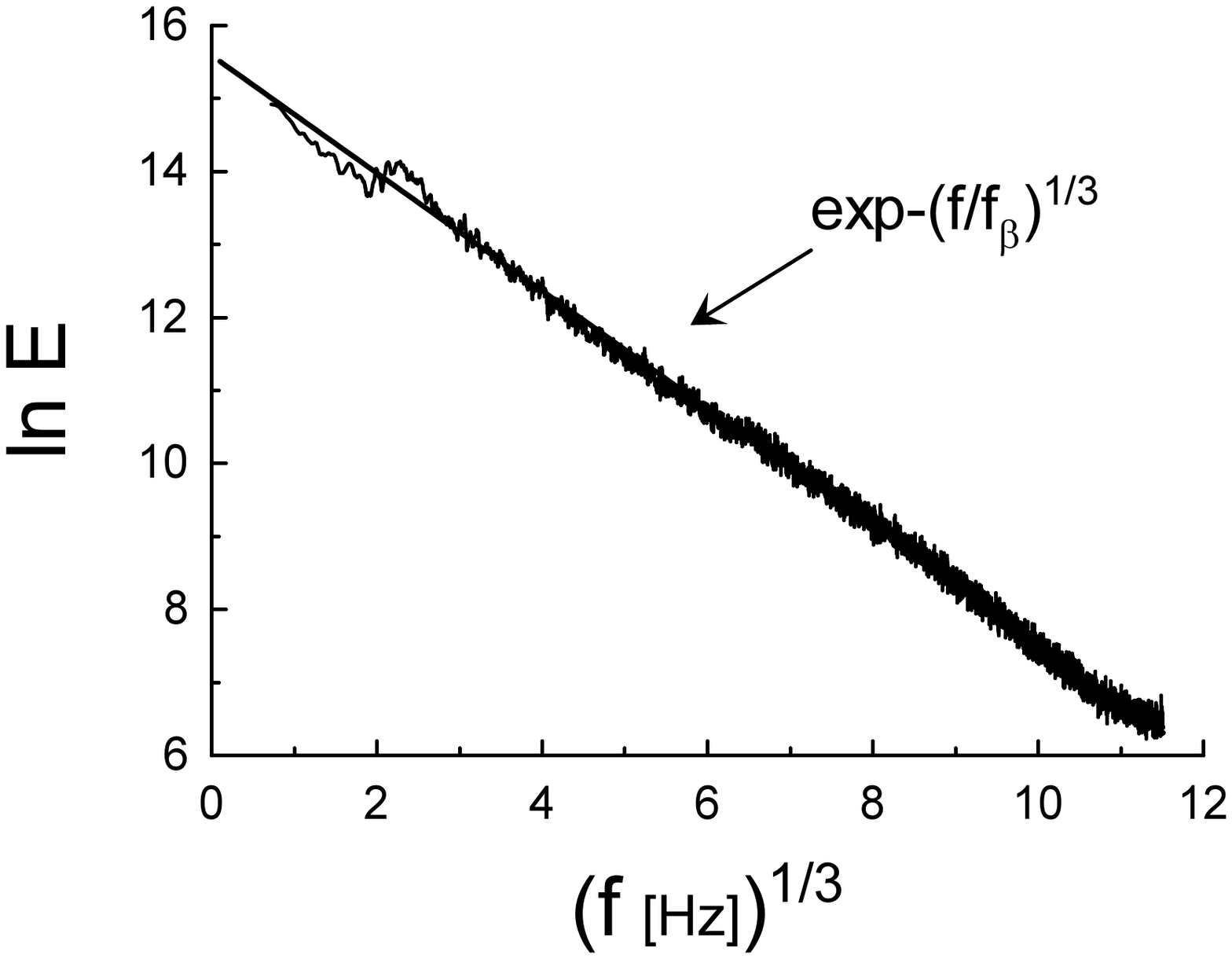}\vspace{-3.8cm}
\caption{\label{fig2} Logarithm of the frequency power spectrum of the temperature fluctuations, measured away from the wake centerline, against $f^{1/3}$.} 
\end{center}
\end{figure}
\begin{figure}
\begin{center}
\includegraphics[width=8cm \vspace{-1.45cm}]{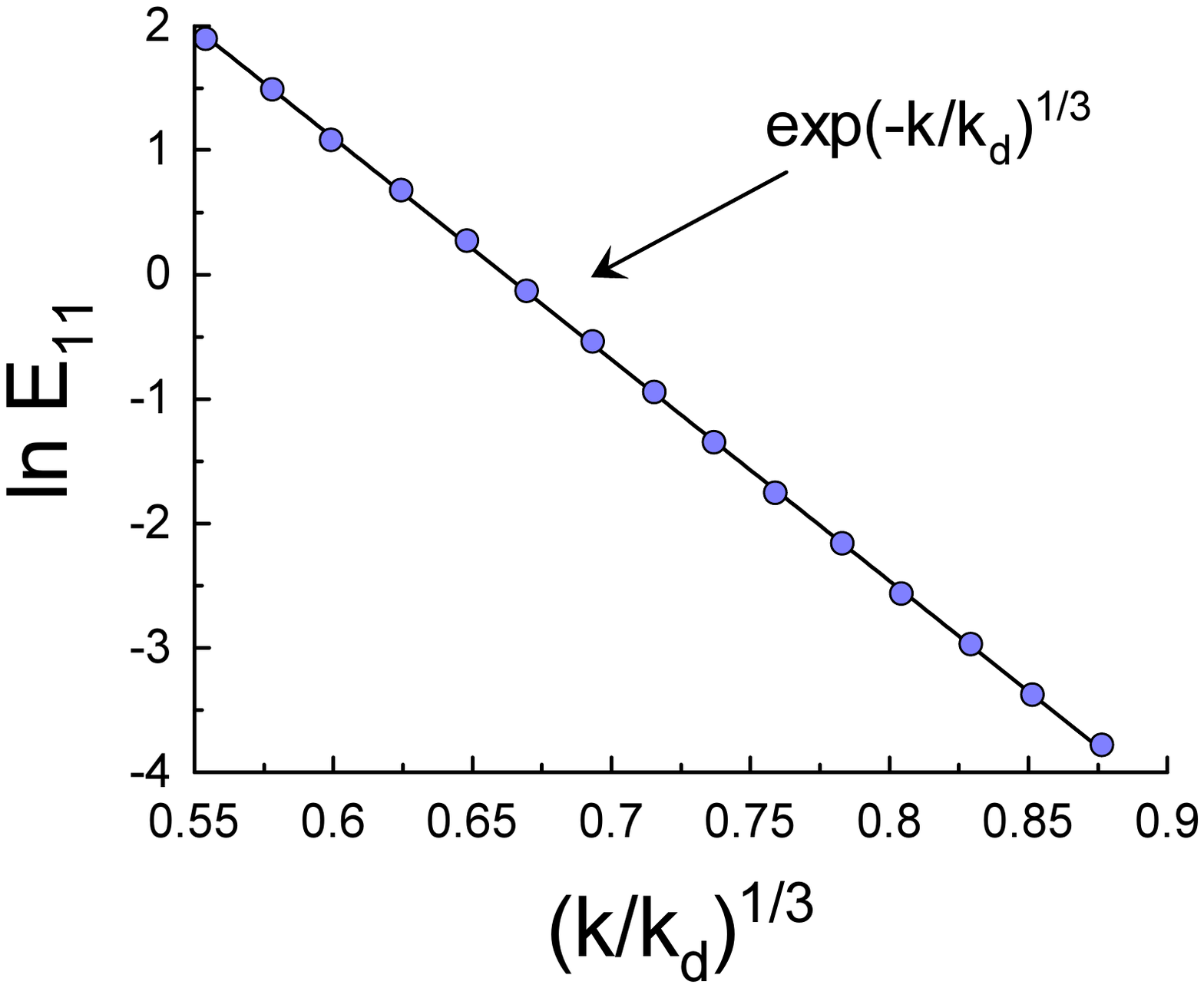}\vspace{-4cm}
\caption{\label{fig3} Logarithm of normalised longitudinal energy spectrum against $(k/k_d)^{1/3}$ in the {\it non-classical} dissipation region of the distances from the grid ($k_d$ is the Kolmogorov's scale). The only chaotic part of the spectrum is shown (corresponding to the insert in the Fig 5.4b of the Ref. \cite{v}). }
\end{center}
\end{figure}
\begin{figure}
\begin{center}
\includegraphics[width=8cm \vspace{-1cm}]{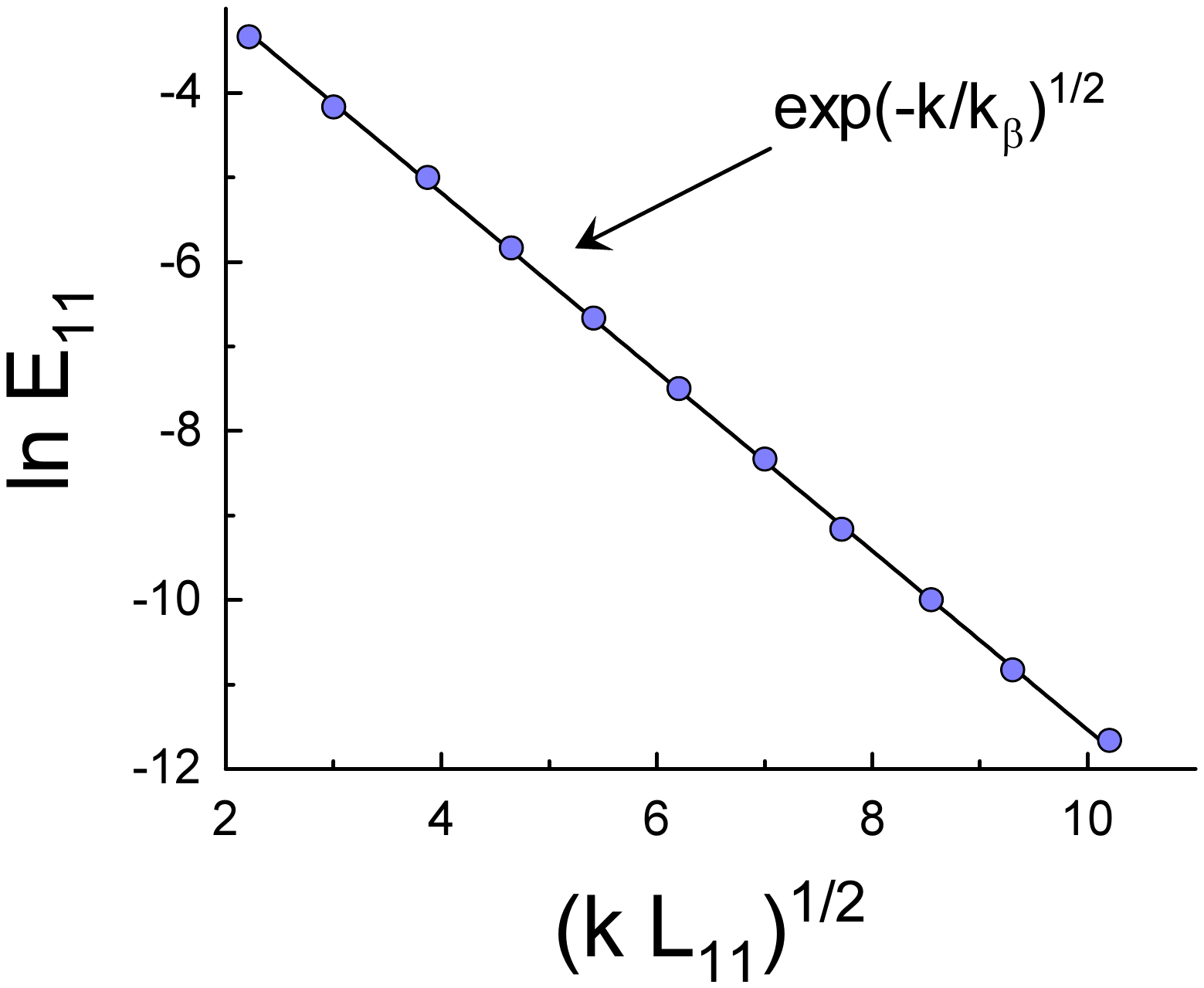}\vspace{-4cm}
\caption{\label{fig4}  Logarithm of normalised longitudinal energy spectrum against $(kL_{11})^{1/2}$ in the {\it classical} dissipation region of the distances from the grid. The only chaotic part of the spectrum is shown} 
\end{center}
\end{figure}

   However, a hard scenario for the spontaneous symmetry breaking can also take place. Indeed, if 
the translational (homogeneity) and rotational (isotropy) space symmetries are broken, and for some reason the soft scenario cannot be realized, the fluid motion has only one of the space fundamental symmetries remained - the Lagrangian relabeling  symmetry \cite{mas}. In this case control on the distributed chaos is switched to the relabeling symmetry.

    Due to the Noether's theorem conservation of helicity in inviscid fluids is a result of the relabeling symmetry  \cite{mor},\cite{y},\cite{pm},\cite{fs}. Therefore the helicity correlation integral (Levich-Tsinober invariant \cite{lt},\cite{fl},\cite{l}) 
$$   
I = \int  \langle  h ({\bf x},t) \cdot   h ({\bf x} + {\bf r},t) \rangle d{\bf r}  \eqno{(5)}
$$  
should be used in the scaling relation for the waves driving the distributed chaos, instead of the vorticity correlation integral (cf Eq. (1)),   
$$
\upsilon (\kappa ) \propto ~I^{1/4}~\kappa^{1/4} \eqno{(6)}
$$
i.e. $\alpha = 1/4$ in this case. Substituting this value of $\alpha$ in the Eq. (4) one obtains
$\beta =1/3$.

\section{Turbulent wake behind cylinder}

In paper Ref. \cite{kss} an experiment with a turbulent wake behind a slightly heated circular cylinder has been described. The experiment was performed in a wind tunnel with Reynolds number $Re_{\lambda} = 130$. The temperature fluctuations (as a passive scalar) were measured in this experiment with a cold-wire probe. Figure 1 shows power spectrum of the temperature fluctuations measured at the farthest downstream location (about 160 cylinder's radii) on the wake centerline. The scales in this figure are chosen to show (as a straight line) correspondence of the data to the spectrum Eq. (3) with $\beta=1/2$ (the soft scenario). Figure 2 shows power spectrum of the temperature fluctuations measured away from the centerline (at distance of about three cylinder's radii). 
The scales in this figure are chosen to show correspondence of the data to the spectrum Eq. (3) with $\beta=1/3$ (the hard scenario). \\

One can see that both the soft and hard scenarios are realized in the wake depending on conditions.

\section{Wakes behind grids}

   Wakes behind grids are widely used in experiments in order to generate 'homogeneous' turbulence. Of course, the turbulence in the wakes of the grids is not homogeneous, especially near the grids (for small $R_{\lambda}$ it is rather complex \cite{b2}). However, these wakes can be successfully used for studying the spontaneous translational symmetry breaking. \\
   
   In series of recent papers Ref. \cite{vv} detailed 
experimental studies of the wakes behind different (active and passive) grids were reported. Here we will use the data obtained in these experiments and taken from Ref. \cite{v}. We will be interested in the regular (passive) grids named in the Refs. \cite{vv},\cite{v} as RG60 (a bi-planar intermediate-blockage square-mesh grid built from rectangular section bars) and RG230 (a mono-planar low-blockage  grid similar to the RG60). The authors of the Refs. \cite{vv}\cite{v} used the Taylor estimate for the kinetic energy dissipation rate
$
\varepsilon \simeq C_{\varepsilon} u^3/L, 
$
with $u$ as the streamwise r.m.s. velocity and $L$ as the longitudinal
integral length-scale, in order to distinguish between the near and far wakes behind the grids. 
They called the (near) wake with non-constant $C_{\varepsilon}$ as non-classical dissipation region. Outside this region $C_{\varepsilon}$ is approximately independent on the Reynolds number in their experiments.

  Figure 3 shows normalised longitudinal energy spectrum for two positions in the {\it non-classical} dissipation region of the distances from the grid (the data were taken from the insert in the Fig. 5.4b of the Ref. \cite{v}, the grid RG230, $R_{\lambda} =300$ and $384$). The scales in this figure are chosen to show (as the straight line) correspondence of the data to the spectrum Eq. (3) with $\beta=1/3$ (the hard scenario). Figure 4 shows normalised longitudinal energy spectrum in the {\it classical} dissipation region of the distances from the grid (the data were taken from the Fig. 5.2b of the Ref. \cite{v}, the grid RG60, $R_{\lambda} =120$). The scales in this figure are chosen to show (as the straight line) correspondence of the data to the spectrum Eq. (3) with $\beta=1/2$ (the soft scenario).  \\
  
  In a recent paper Ref. \cite{kro} results of an experiment with multi-scale grids were reported. 
The large recirculating wind tunnel was used for this experiment (see for details Ref. \cite{kd}).  The Reynolds number based on $M_1~$ $Re_M =6.0 \times 10^4$ (where $M_1$ is the largest mesh size for the multi-scale grid). The multi-scale grids inject energy in the flow at a large number
of scales. In particular, at present experiment the geometric scales of the multi-scale grid are distributed about over the whole range of the turbulent flow spectrum. 
\begin{figure}
\begin{center}
\includegraphics[width=8cm \vspace{-0.8cm}]{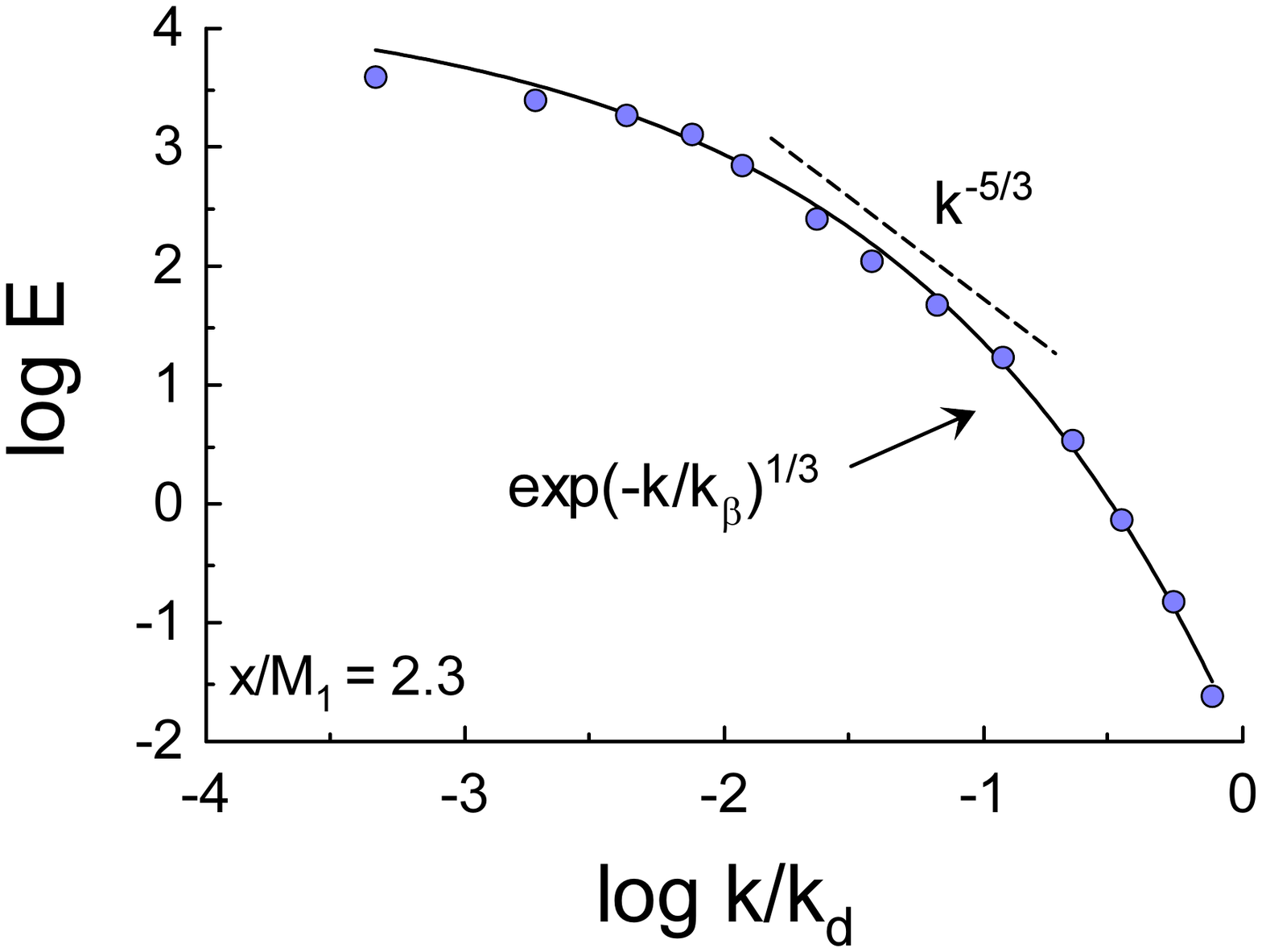}\vspace{-3.9cm}
\caption{\label{fig5} Longitudinal energy spectrum near the multi-scale grid: $x/M_1 =2.3$. The solid curve indicates the hard scenario.}
\end{center}
\end{figure}

\begin{figure}
\begin{center}
\includegraphics[width=8cm \vspace{-1cm}]{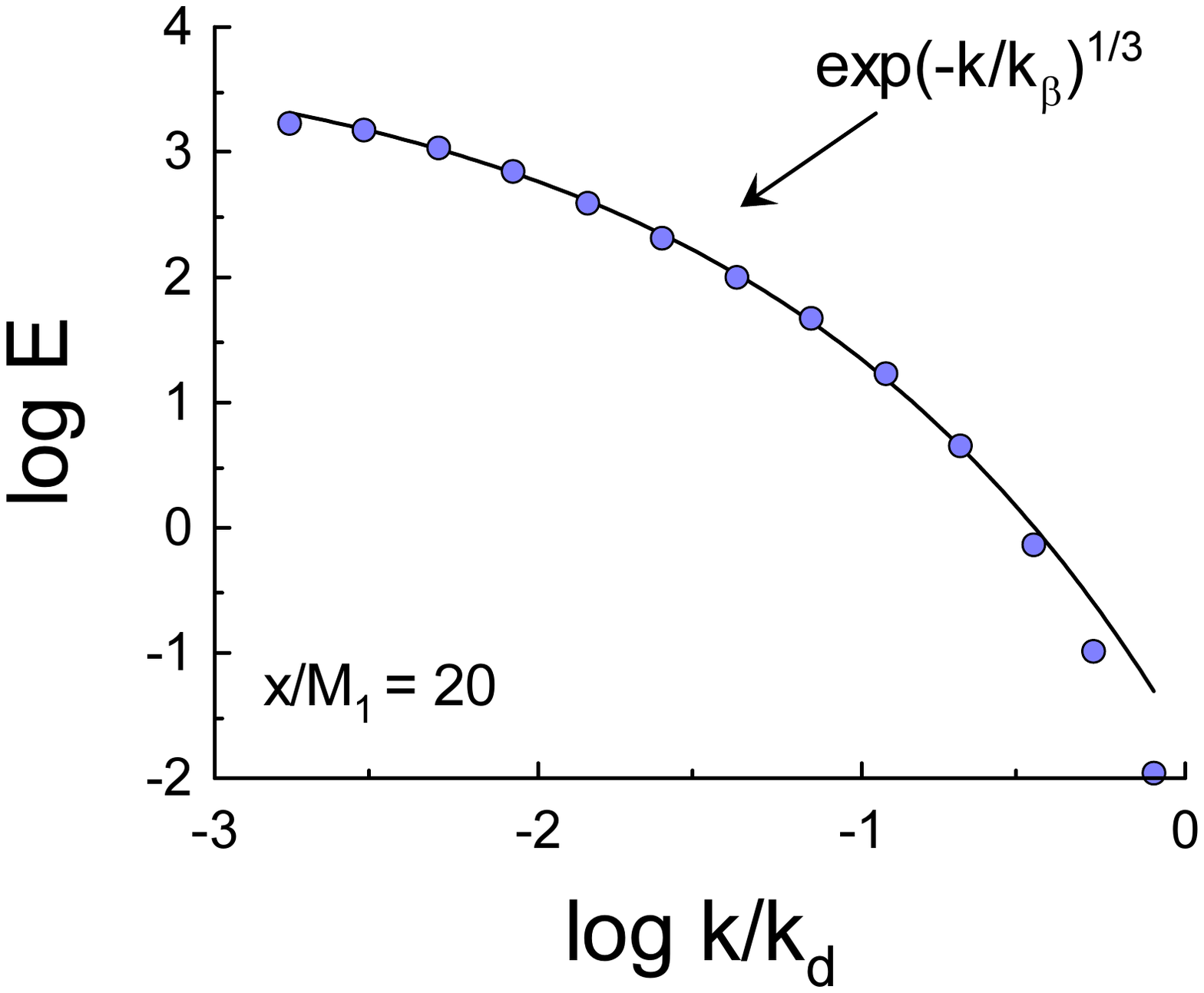}\vspace{-3.8cm}
\caption{\label{fig6} Longitudinal energy spectrum at distance $x/M_1 =20$ from the multi-scale grid. The solid curve indicates the hard scenario at small $k/k_d$.}
\end{center}
\end{figure}

\begin{figure}
\begin{center}
\includegraphics[width=8cm \vspace{-1cm}]{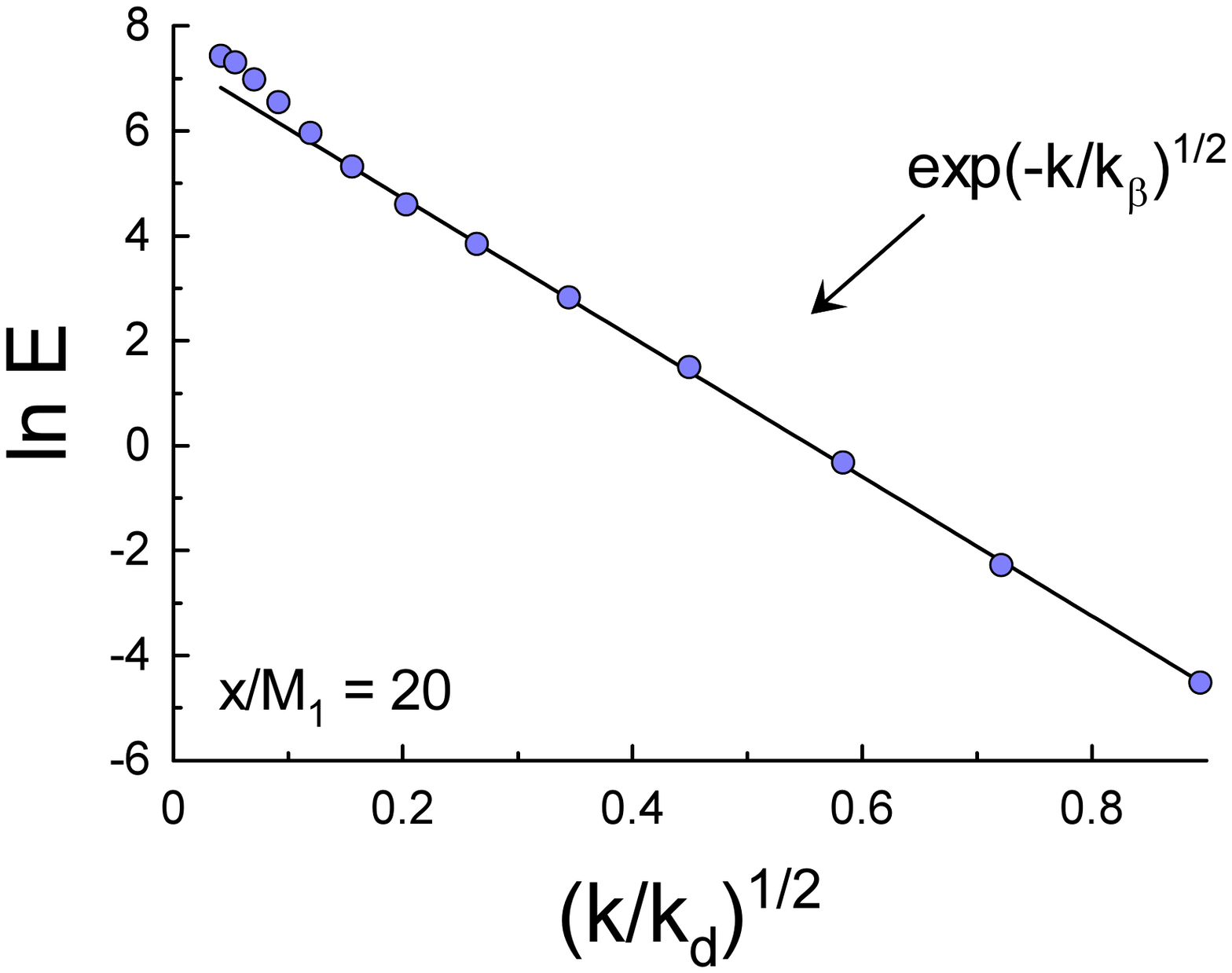}\vspace{-3.5cm}
\caption{\label{fig7} The same as in Fig. 6 but in the scales chosen to show (as the straight line) correspondence of the data to the spectrum Eq. (3) with $\beta=1/2$ (the soft scenario at large $k/k_d$).}
\end{center}
\end{figure}
   Figure 5 shows (in the log-log scales) longitudinal energy spectrum near the multi-scale grid: $x/M_1 =2.3$ (the data correspond to Fig. 9 of the Ref. \cite{kro}). The solid curve (the best fit) is drawn in order to indicate correspondence of the data to the spectrum Eq. (3) with $\beta=1/3$ (the hard scenario). Unlike the above described situation (Fig. 3) the spectrum Eq. (3) with $\beta =1/3$ covers about entire spectrum in this case. The dashed straight line with the slope -5/3 indicates a hint of the Kolmogorov scaling law.\\
   
   Figure 6 shows analogous spectrum at distance $x/M_1 = 20$ from the grid (the data correspond to Fig. 8 of the Ref. \cite{kro}). It is interesting that at this distance the spectrum Eq. (3) with $\beta =1/3$ covers a range corresponding to comparatively small values of $k/k_d$. Figure 7 shows the same data as in Fig. 6 but in the scales chosen to show (as the straight line) correspondence of the data to the spectrum Eq. (3) with $\beta=1/2$ (the soft scenario). At the large $k/k_d$ and $x/M_1$ turbulence is more homogeneous hence the soft scenario.\\
  
  As in the previous section both the soft and hard scenarios are realized in the grid wakes depending on conditions.
 
\section{Wakes behind grids in superfluid}

A liquid phase of $~^4$He is called He I with Navier-Stokes dynamics above $T_{\lambda} \simeq 2.17$ K, and He II (superfluid) below $T_{\lambda}$. The two-fluid model can be used for this superfluid composed of a Navier-Stokes component and a superfluid component with quantized vorticity and zero viscosity. The ratio $\rho_{s}/\rho=0$ at $T_{\lambda}$ and increases to 1 with $T \rightarrow 0$ K ($\rho_s$ is density of the superfluid component and $\rho$ is total density).

  In a recent experiment reported in Ref. \cite{sal} a wake behind a grid was studied above and below $T_{\lambda}$.  In He I the grid mesh size Reynolds number $R_M \simeq 10^5 \sim 2\times 10^6$, in He II $R_M \simeq 1.5 \times 10^4 \sim 2 \times 10^5$ (in the last case the quantum of circulation was used instead of viscosity in defining the Reynolds number).
  
  Figure 8 shows velocity spectra obtained in this experiment at temperatures below $T_{\lambda}$ at distance $x/M \simeq 138$ downstream the grid (the data were taken from Fig. 4 of the Ref. \cite{sal} and only relevant peaks are shown). The dashed lines (the best fit) indicate the stretched exponential spectrum Eq. (3) with $\beta = 1/3$ (the hard scenario) for the both observed spectra. 
  
    An interesting hint of a scaling (of Kolmogorov's type with $E \propto f^{-5/3}$) has been shown as a straight solid line for the upper spectrum (2.1 K, 3.3 m/s) in Fig. 8 (cf Fig. 5). This hint is discussed in detail in the Ref. \cite{sal} and has been also supported there by comparison with the Kolmogorov constant. Usually the range of the distributed chaos is not overlapped considerably with the scaling region because the exponential spectra are typical for smooth (sub)systems whereas the scaling spectra are typical for the rough ones. For $T < T_{\lambda}$ one could relate the overlapping to the two-fluid situation. But the problem with this explanation is that the spectral picture observed in this experiment for $T = 2.6~$K $> T_{\lambda}$ (3.3 m/s) is very similar to that presented in the Fig. 8.

\begin{figure}
\begin{center}
\includegraphics[width=8cm \vspace{-1cm}]{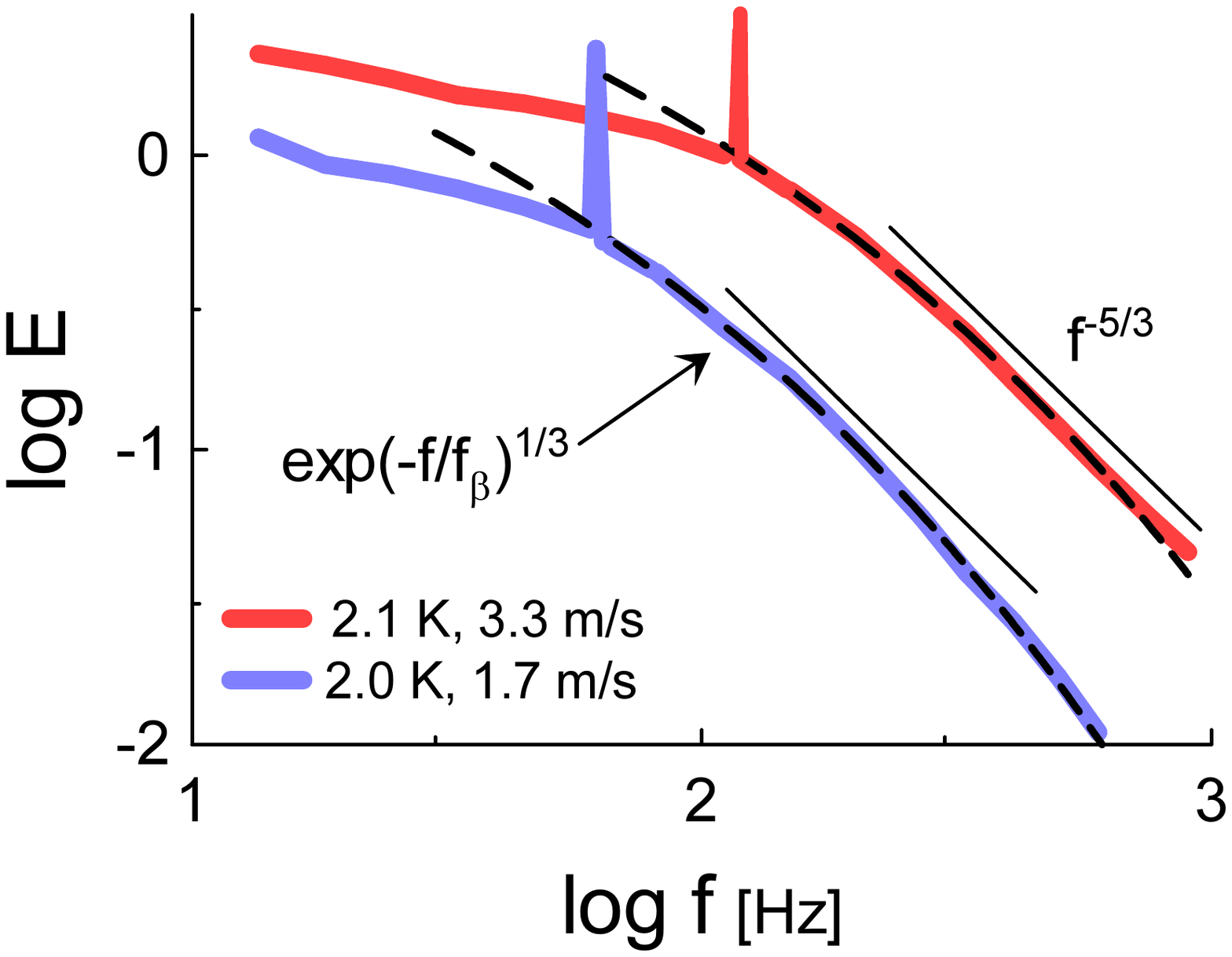}\vspace{-4cm}
\caption{\label{fig8} Power spectra of velocity behind a grid in a superfluid (log-log scales, $E$ is in arbitrary units).}
\end{center}
\end{figure}

\begin{figure}
\begin{center}
\includegraphics[width=8cm \vspace{-1.25cm}]{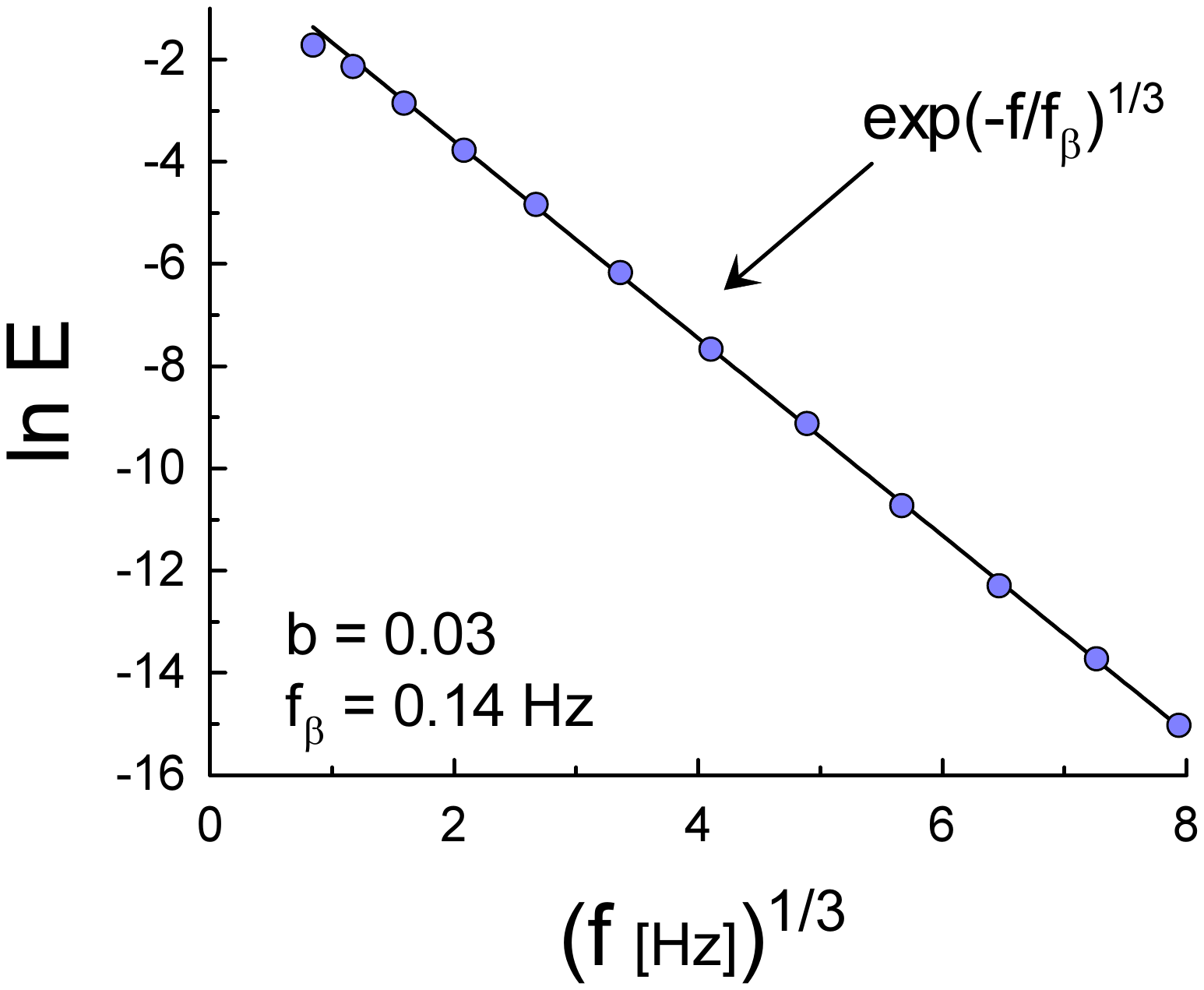}\vspace{-3.5cm}
\caption{\label{fig9} Logarithm of normalised power spectrum of vertical component of the water velocity fluctuations for $b=0.03$}
\end{center}
\end{figure}

\begin{figure}
\begin{center}
\includegraphics[width=8cm \vspace{-1.2cm}]{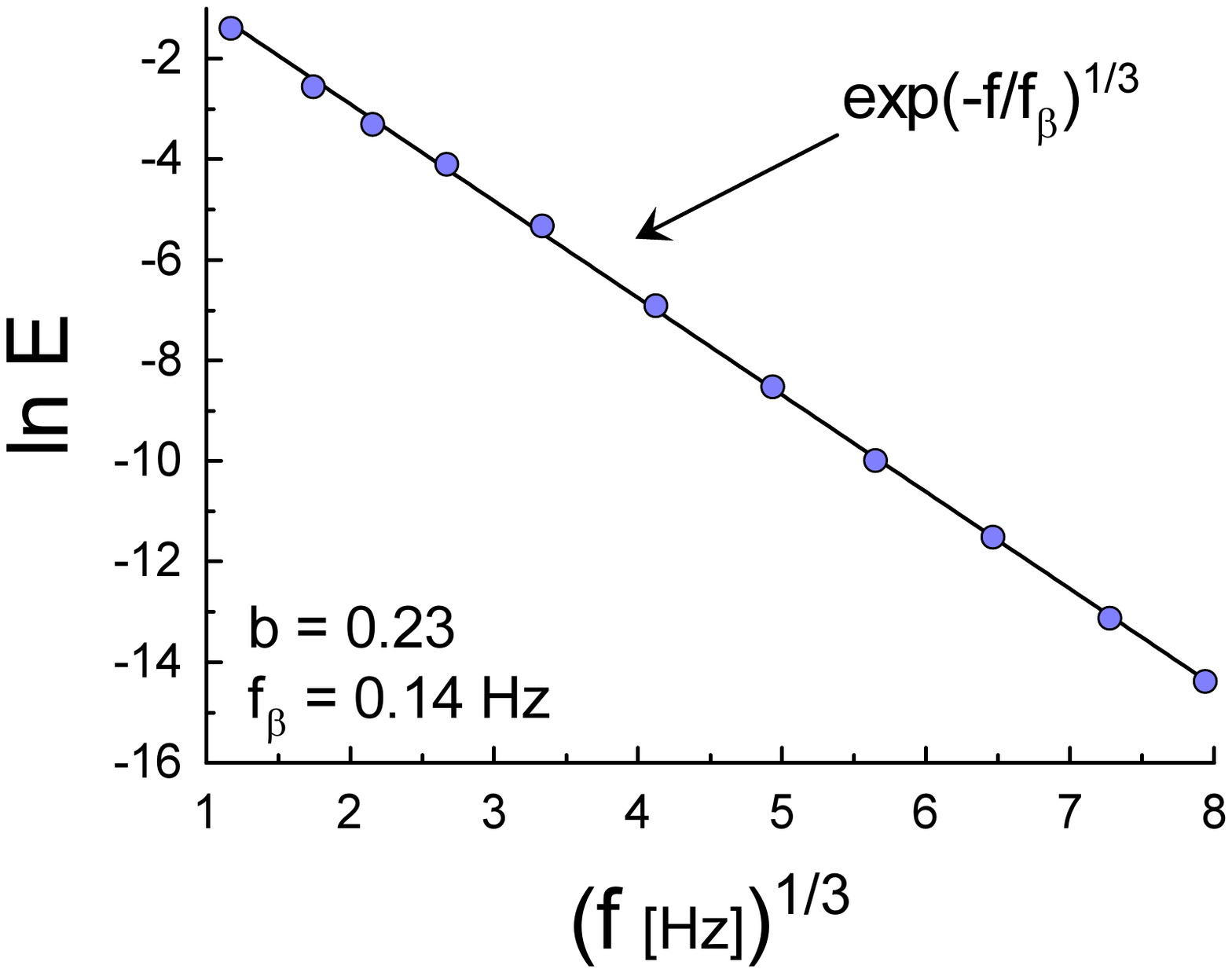}\vspace{-3.8cm}
\caption{\label{fig10} The same as in Fig. 9 but for $b=0.23$.}
\end{center}
\end{figure}
\section{Turbulent bubbly flows}

  The bubble wakes generate turbulence in surrounding liquid. It is known that even a tiny concentration of bubbles changes the velocity spectra cardinally \cite{twt}. It is typical for spontaneous symmetry breaking. 
  
  To quantify the phenomenon the  'bubblance' parameter was introduced in Ref. \cite{lb} (see also Refs. \cite{twt} and \cite{rll})
$$
b = \frac{1}{2} \frac{\alpha U_r^2}{u_0^2}  \eqno{(7)}
$$
with  $U_r$ as the bubble rise velocity in
still liquid, $u_0$ as the typical velocity fluctuation in the liquid without of bubbles, and with $\alpha$ as concentration of the bubbles. This is a 'kinetic energy ratio' parameter and it can be relevant to the energy spectral analysis. The value $b=0$ corresponds to single-phase situation (without bubbles). We naturally will be interested here in small values of the parameter $b$ relevant to the spontaneous symmetry breaking. 

   In recent paper Ref. \cite{twt} the data obtained in an experiment with a wide range of the parameter $b$ were presented (we will use these data obtained for small values of $b$). In this experiment an 8 m high vertical water tunnel, with an active grid generating coflowing turbulent upward bubbly flow, was used ($R_{\lambda} =170$ for $b=0$) .  
   
   Figure 9 and 10 show normalised power spectrum of vertical component of the water velocity fluctuations for $b=0.03$ and $0.23$ respectively (the data were taken from Fig.6f of the Ref. \cite{twt}). The scales in these figures are chosen to show (as the straight line) correspondence of the data to the spectrum Eq. (3) with $\beta=1/3$ (the hard scenario). It is worth noting that value of the parameter $f_{\beta} \simeq 0.14$ Hz is the same for these two cases (that can mean a tuning of the distributed chaos to this low-frequency value at the hard spontaneous symmetry breaking). \\

I thank K. R. Sreenivasan and P.-A. Krogstad for sharing their data, D. Lohse for attracting my attention to the experimental studies of his group and A. Pikovsky for explanations.

\end{document}